\documentclass[10pt,journal]{IEEEtran}

\ifCLASSOPTIONcompsoc

\usepackage[nocompress]{cite}
\else

\usepackage{cite}
\fi

\ifCLASSINFOpdf

\else

\fi

\usepackage{subfigure}
\usepackage{graphicx}
\usepackage{balance}
\usepackage{amsmath,amsthm, amssymb}
\usepackage{amssymb,wasysym}
\usepackage{amsfonts,balance}
\usepackage{mathrsfs}
\usepackage{cite}
\usepackage{color}
\usepackage{url}
\usepackage{bm}
\usepackage{wrapfig}
\usepackage{verbatim}
\usepackage{dsfont}
\usepackage{amsmath}
\usepackage{algorithm}
\usepackage{algpseudocode}
\usepackage{xurl}
\usepackage{tabularx}
\usepackage{enumitem}
\usepackage[colorlinks=true, linkcolor=black, citecolor=black, urlcolor=black, bookmarks=false]{hyperref}

\theoremstyle{definition}

\allowdisplaybreaks[3]

\ifodd 1
\else

\fi

\graphicspath{{figures/}}

\begin{document}
	\title{Space-ground Fluid AI for 6G Edge Intelligence}
	
	\author{Qian Chen, Zhanwei Wang, Xianhao Chen, Juan Wen, Di Zhou, Sijing Ji, Min Sheng, and Kaibin Huang
		
		\thanks{

			Q. Chen, Z. Wang, X. Chen, and K. Huang are with the Department of Electrical and Electronic Engineering, The University of Hong Kong, Hong Kong. 

            J. Wen, D. Zhou, S. Ji, and M. Sheng are with the State Key
Laboratory of Integrated Service Networks, Xidian University, Xi'an 710071,
Shaanxi, China.

            Corresponding author: K. Huang (Email: huangkb@eee.hku.hk).
			
		}
	}
	
	\markboth{} {Shell \MakeLowercase{\textit{et al.}}: Bare Demo of IEEEtran.cls for IEEE Journals}

\IEEEtitleabstractindextext{	\begin{abstract}
 Edge artificial intelligence (AI) and space-ground integrated networks (SGINs) are two main usage scenarios of the sixth-generation (6G) mobile networks. Edge AI supports pervasive low-latency AI services to users, whereas SGINs provide digital services to spatial, aerial, maritime, and ground users. This article advocates the integration of the two technologies by extending edge AI to space, thereby delivering AI services to every corner of the planet. Beyond a simple combination, our novel framework, called space-ground fluid AI, leverages the predictive mobility of satellites to facilitate fluid horizontal and vertical task/model migration in the networks. This ensures non-disruptive AI service provisioning in spite of the high mobility of satellite servers. The aim of the article is to introduce the (space-ground) fluid AI technology. First, we outline the network architecture and unique characteristics of fluid AI. Then, we delve into three key components of fluid AI, i.e., fluid learning, fluid inference, and fluid model downloading. They share the common feature of coping with satellite mobility via inter-satellite and space-ground cooperation to support AI services. Finally, we discuss the considerations for the real-world deployment of fluid AI and identify further research opportunities. \\
	\begin{IEEEkeywords}
    AI techniques, edge intelligence, satellite networks, mobility, adaptive scheduling. \end{IEEEkeywords}
	\end{abstract} }
	\maketitle	
	\IEEEdisplaynontitleabstractindextext
    
    \IEEEpeerreviewmaketitle
	
	\section{Introduction}
The commercialization of sixth-generation (6G) mobile networks is expected by 2030. According to the International Telecommunication Union (ITU), two of three new usage scenarios in 6G mobile networks—as compared with fifth-generation (5G) mobile networks—are ``integrated artificial intelligence (AI) and communication" and “ubiquitous connectivity". For the former scenario, edge AI has emerged as a mainstream direction for 6G development~\cite{Letaief2022JSAC}. By supporting distributed machine learning, AI inference, and AI model delivery at the network edge, 6G mobile networks can enable a wide range of applications, including extended reality, autonomous driving, and collaborative robots. For the latter scenario, 6G is anticipated to form space–ground integrated networks (SGINs) that seamlessly integrate terrestrial and non-terrestrial (satellite) networks~\cite{huaweiwhitepaper2022}. In this context, numerous countries, as well as leading technology companies such as SpaceX, have joined the race to launch mega-constellations of interconnected satellites as a new type of telecommunication infrastructure.

Since modern satellites are equipped with substantial computing resources (e.g., equivalent to 60+ Linux computers for a Starlink satellite~\cite{Space-Linux}), they not only serve as communication nodes but also function as computing servers. Prior studies have explored the use of satellites as mobile edge computing (MEC) servers, where satellites process computing tasks offloaded from ground users, thereby alleviating the computing workload of ground users. To optimize task offloading and resource allocation, various optimization-based techniques (e.g., convex optimization, game theory, and heuristic algorithms)~\cite{9978924} and AI-driven methods (e.g., reinforcement learning and deep learning) have been developed~\cite{8672604}. 
In very recent years, with the growing dominance of AI-related computation, researchers have specifically considered AI-capable satellites functioning as edge AI servers for edge learning \cite{9982621} and edge inference \cite{10694785}. Research works in this area involve tailored design for training/inference accuracy maximization or neural network partitioning. However, the integration of edge AI and SGINs is still in its nascent stage.

Embracing the aforementioned two trends of 6G, this article advocates the integration of edge AI and SGIN technologies to facilitate the extension of AI services from ground to space. Connecting tens of thousands of these computing-enabled orbiting satellites to terrestrial servers will create an expansive three-dimensional (3D) fluid computing platform available anywhere and anytime. This paradigm not only bridges the digital divide by delivering AI services to every corner of the earth but also supports mission-critical intelligent applications, such as AI-empowered search and rescue, in remote and disaster-stricken areas. However, the rapidly changing satellite network topologies and limited communication rates of space–ground links present significant challenges:
	\begin{itemize}
		\item \textbf{Ensuring service continuity under high satellite mobility:} The first critical challenge is how to manage task migration effectively to ensure service continuity in the context of the high-speed movement of satellites, which fundamentally differ from terrestrial edge servers with fixed locations. As such, inter-node handovers are frequent for networks that include low-Earth-orbit (LEO) satellites, occurring every few minutes and leading to significant difficulty in ensuring service continuity~\cite{9422812}.

		\item \textbf{Overcoming the communication bottleneck in feeder links:} Given the severe signal attenuation and excessively long propagation delay between space and ground (the average round-trip time is 50 ms when connecting LEO satellites~\cite{RTT}), it is extremely challenging to support low-latency and data-intensive edge AI services.
        
	\end{itemize}

To tackle the above challenges, we propose a way to seamlessly integrate edge AI with SGINs to extend the two-dimensional (2D) edge-AI architecture into space, leading to a new framework named space–ground fluid AI. This name is inspired by the fact that water can fit into a container with an arbitrary shape. With high mobility, the network topology of an SGIN is analogous to a flexible, ever-changing container, whereas the network traffic flow—that is, the AI model parameters and data features—behaves like water, continuously flowing both horizontally (across space) and vertically (between space and ground) in response to the evolving container. Although the dynamic shape of the ``container” causes considerable difficulty, researchers must develop new fluid AI techniques to adapt ``water” according to ``containers”, thereby ensuring robustness and continuity of services. Underpinning the fluid AI technologies are satellites’ predictable and repeatable movement patterns, which differentiate such networks from their static ground or randomly moving unmanned aerial vehicle (UAV) counterparts. Note that the predictable mobility of edge servers has never been factored into traditional resource-management schemes for edge AI, calling for new designs targeting fluid AI in the space.

The proposed fluid AI framework comprises three core AI techniques: fluid learning, fluid inference, and fluid model downloading schemes. First, to overcome the long model training time required for an SGIN, we design an infrastructure-free ``model-dispersal” federated learning (FL) scheme for fluid AI, which accelerates training convergence by mixing model parameters across regions via orbiting satellites—analogous to seed dispersal via animals. In this way, satellite mobility is transformed from a difficulty into an asset. Second, by considering the heterogeneous computing capabilities of satellite servers, we design an inference framework for fluid AI that cascades model blocks distributed over satellites. The control of the propagation depth in the resultant global model introduces a tradeoff between inference accuracy and latency. In addition, the strategies of horizontal and vertical model migration ensure task continuity. Finally, to enable efficient model downloading to ground users, we propose a parameter-sharing caching scheme for storing models on satellites and design a parameter-sharing multicasting scheme to overcome the low data rate between space and ground.

The remainder of this article is organized as follows. We first present the network architecture and describe the salient characteristics of fluid AI in Section~\ref{sec:overview}. Then, we illustrate three fluid AI techniques—namely, fluid learning, fluid inference, and fluid model downloading—in Section~\ref{section:key_techniques}. Deployment considerations and future directions of fluid AI are outlined in Section~\ref{sec:deploy_future}, followed by concluding remarks in Section \ref{sec:conclusion}.

 \section{Overview of Space-ground Fluid AI}\label{sec:overview}
In this section, we first present the network architecture of SGINs. Following this, we highlight research opportunities for AI design arising from the unique characteristics of SGINs. This section serves as the foundation and motivation for the techniques we propose in Section \ref{section:key_techniques}.

 \subsection{Network Architecture of Fluid AI}
\begin{figure}[h]
 	\centering
 	\includegraphics[width = 0.4\textwidth]{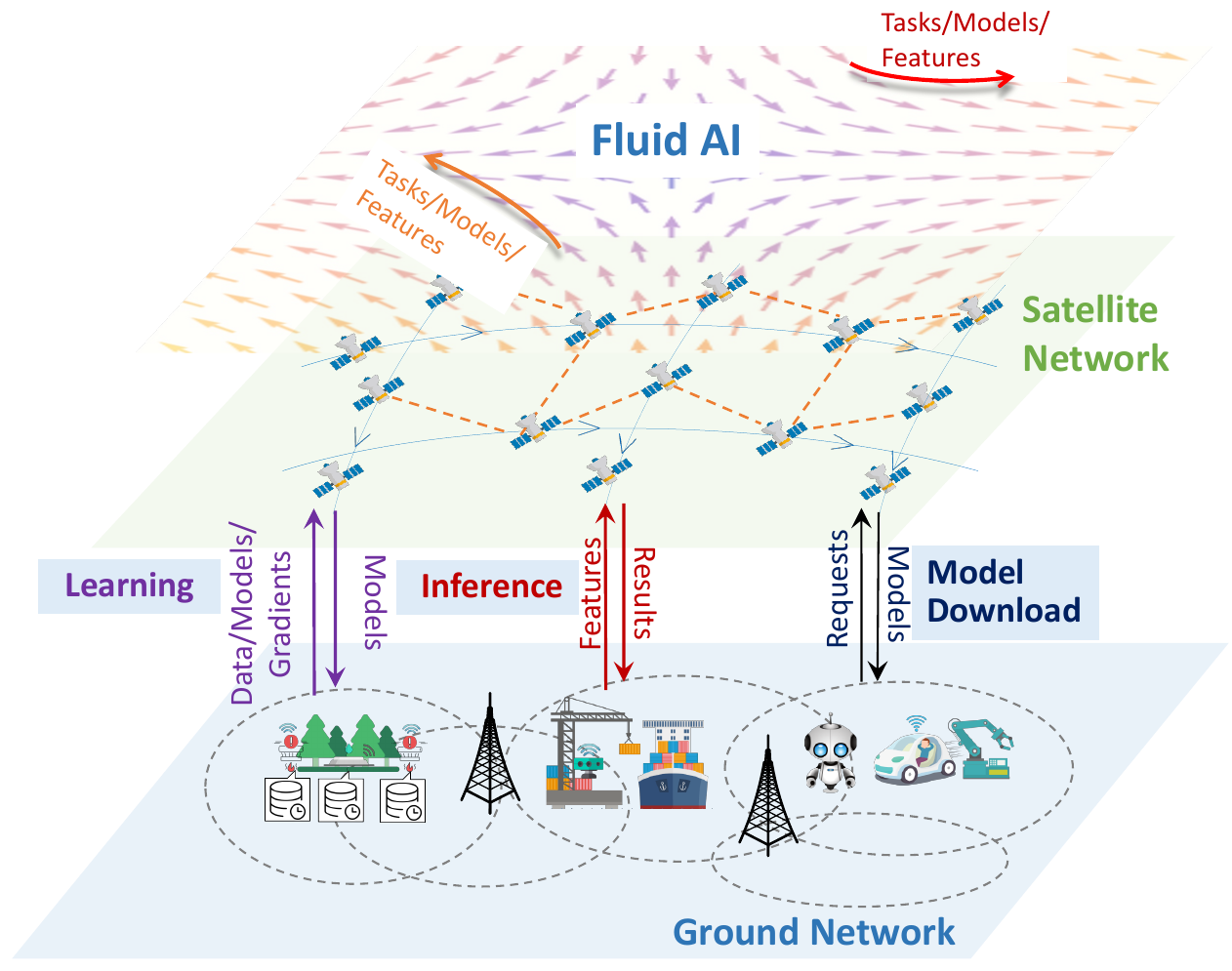}
 	\caption{Distributed learning,
 		inference, and model downloading in fluid AI systems. \label{fig:SG-FAI_system}}
 \end{figure}

 A multi-user fluid AI system is depicted in Fig.~\ref{fig:SG-FAI_system}, made up of three main components.

1) \textit{Ground Internet of Things (IoT) devices}: These devices are distributed across various geographical locations, each holding a small portion of heterogeneous datasets and periodically generating different tasks.

2) \textit{AI-empowered satellites}: The satellites follow predetermined trajectories and possess more powerful processing capabilities than the ground clients. They can perform model training and task inference for ground users and serve as edge caches for the AI model library.

3) \textit{Ground stations}: The ground stations connect to the satellites via feeder links and act as the control centers of the fluid AI system. Since they are linked to data centers through wired optical cables, these ground stations can effectively function as the cloud. They are responsible for calculating efficient resource-management strategies and directing the satellites and ground devices to execute the instructions. Additionally, with their powerful computing capabilities, the ground stations can cooperate with the satellites to provide cloud-edge computing.

Such a fluid AI system serves as a platform for implementing distributed learning, inference, and model downloading within SGINs based on the enabling technologies described below.

  \subsection{Characteristics of Fluid AI}\label{subsec:opportunity}
Compared with terrestrial networks, fluid AI possesses unique characteristics that can be taken advantage of to design a more effective framework for enabling space–ground AI provisioning:
  
  \subsubsection{Predictable channel conditions}
The fluid AI technology benefits from the non-causal knowledge of SGINs’ dynamic topology. Such knowledge, called ephemeris information, can be translated into channel predictability, enabling intelligent task/model migrations and resource management and thereby improving system performance and efficiency. To achieve this, machine learning-based channel predictors, such as long short-term memory (LSTM) and transformer models, can be developed and trained using practical measurement datasets and non-causal network topology information. The predictable channel conditions in fluid AI networks can support resource and mobility management, as presented in Section \ref{section:key_techniques}, ensuring the continuity of AI services provisioning.
  	
    \subsubsection{Repeated orbital motion}
 We can also take advantage of the repeated orbital motion of satellites for effective resource and mobility management. Over a short period of time, during which the effects of Earth’s rotation can be neglected, satellites within the same orbit can be assumed to follow identical trajectories. In other words, each subsequent satellite follows the path of its predecessor. Even for the time period when Earth’s rotation cannot be neglected, the ground tracks of adjacent orbits in a LEO satellite constellation still repeat over a specific time interval, which is determined by the difference in the right ascension of the ascending node (RAAN) between adjacent orbits. The repeated movement of satellites introduces distinctive opportunities for cross-region task migration and model sharing, which will be elaborated in Section~\ref{section:key_techniques}.

\section{Key Techniques of Space-ground Fluid AI}\label{section:key_techniques}
As previously mentioned, the unique characteristic of fluid AI—in contrast to terrestrial edge AI—lies in leveraging the predictive mobility of satellites to enable intelligent horizontal and vertical task/model migration. We will implement the most advanced communication-computing integration methods, as outlined in the 6G paradigm. The natural integration of AI technologies with radio resource management in fluid AI is intended to achieve continuous service, optimal end-to-end (E2E) performance, and power efficiency. Within this framework, we will focus on developing three key technologies: fluid learning, fluid inference, and fluid model downloading, as depicted in Fig.~\ref{fig:SG-FAI_system}.

\subsection{Space-ground Fluid Learning}\label{subsec:learning}
Large-scale model training with vast amounts of data in remote areas is essential for supporting mission-critical applications such as healthcare, search and rescue, agriculture, and more. However, this goal is hindered by the lack of communication-computing infrastructure. Although satellite networks can address this issue, existing learning schemes in SGINs either have long convergence durations or require costly inter-satellite links (ISLs). To tackle these challenges, we introduce the first component of fluid AI, termed fluid learning, which is designed to facilitate distributed training across large geographical regions by exploiting satellite mobility.
FL is a decentralized machine learning approach widely studied for SGIN systems, in which ground edge devices, such as sensors or mobile phones, periodically upload their updated models or gradients to a satellite for model aggregation. However, due to the limited serving area of an LEO satellite, achieving model consensus on a global scale presents substantial challenges. Existing SGIN FL frameworks for addressing this issue generally fall into two categories: 

 1) In \textit{ground station-assisted hierarchical FL} \cite{10121575}, satellites further forward models to a ground station to perform global aggregation. However, the learning efficiency is severely hampered by the excessive waiting time, as there is likely to be a prolonged delay before a satellite encounters a ground station. 
 
 2) \textit{ISL-assisted decentralized FL} \cite{9982621} enables model exchanges and averaging among satellites through ISLs, thereby circumventing the prolonged waiting time. However, laser terminals are prohibitively expensive (e.g., each costs approximately 500k USD, according to the Space Development Agency contract \cite{laserISL}), so this approach is not universally applicable. For instance, ISLs are not included in phase 1 of Starlink, among many other satellites. Furthermore, even with ISLs available on these satellites, forwarding models via multi-hop ISLs still incur non-negligible bandwidth costs.

\begin{figure*}[ht]
	\centering
	\subfigure[Illustration of the proposed scheme.]{\includegraphics[width =0.3\textwidth]{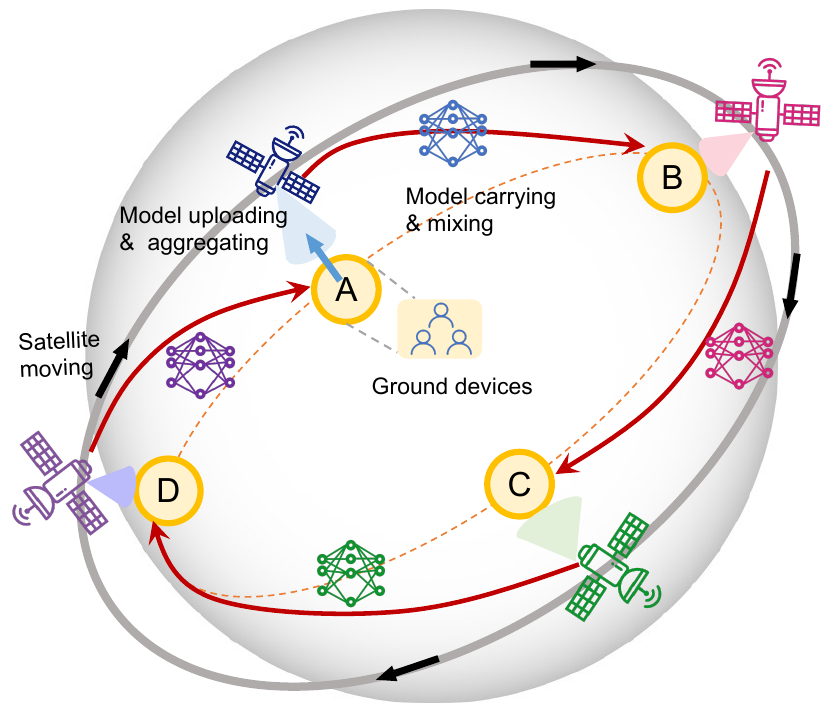}\label{fig:model_dispersal_FL}}
    \quad
	\subfigure[Test accuracy vs. convergence time under different FL schemes on CIFAR-10 (left) and MNIST (right).]{\includegraphics[width =0.5\textwidth]{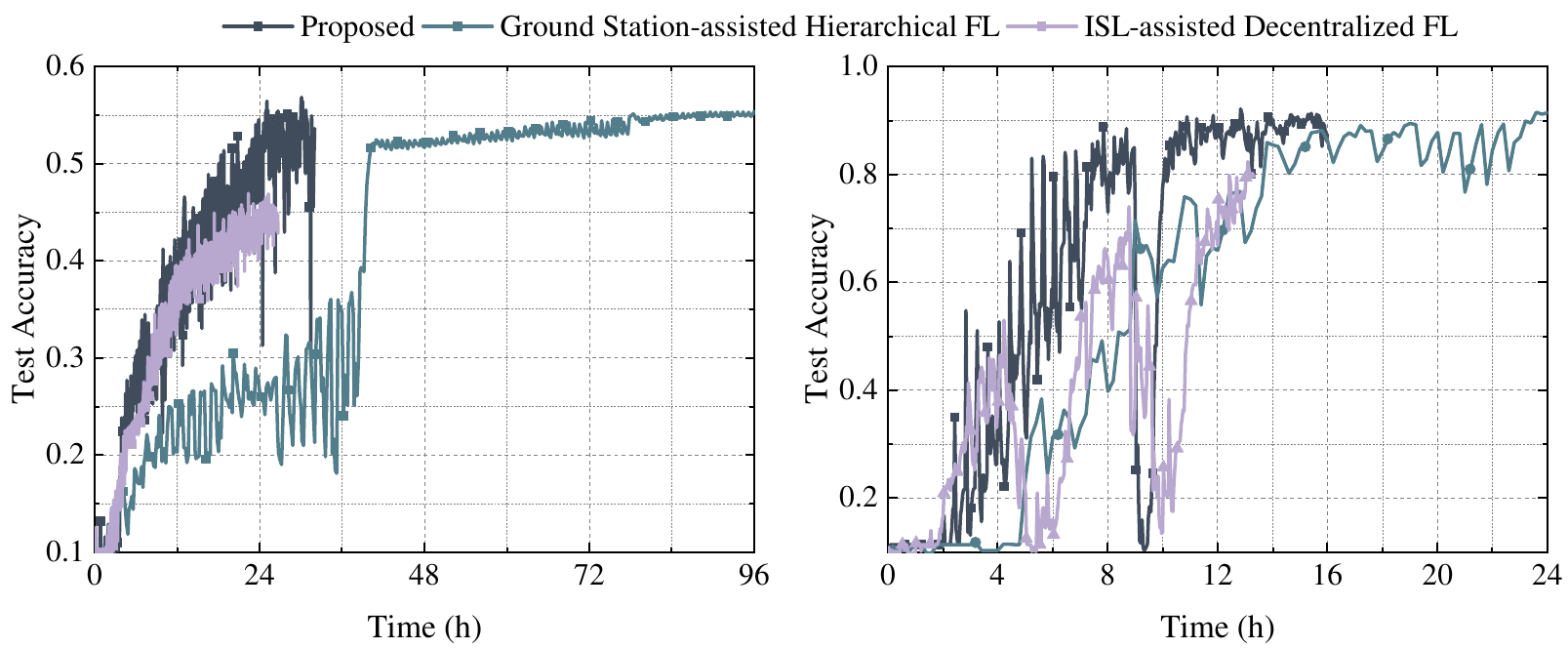}\label{fig:learning_compare}}
	\caption{Infrastructure-free ``model-dispersal" FL scheme.} 
\end{figure*}

Since the above schemes rely on either ground infrastructure or ISLs, it is important to develop a novel infrastructure-free “model-dispersal” FL scheme by exploiting satellite mobility. Thanks to the store-carry-forward capabilities of orbiting satellites, a satellite can carry the area model to the next area for knowledge fusion. This allows for model dispersal across different areas via satellite carriers without requiring the assistance of ISLs or ground stations—analogous to seed dispersal via animals \cite{chen2024fedmeld}. 
In this scheme, the training process within the same cluster (e.g., clusters A, B, C, and D in Fig. \ref{fig:model_dispersal_FL}) follows a synchronous pattern. Across different clusters, the training process follows an asynchronous pattern, as clients may perform training with satellites that do not carry the model from the previous region. This asynchronous approach allows for more flexible training progress and reduces waiting time between different areas. To ensure the robustness of synchronous FL within each cluster, we implement strategies to handle dropout clients. Specifically, once participating clients are selected in each global round, a fixed time window is assigned. If a client fails to upload its model within this specified period due to communication delays or temporary user-to-satellite link unavailability, its update is excluded from aggregation. The aggregation then proceeds with standard algorithms such as weighted averaging.

Fig. \ref{fig:learning_compare} compares the test accuracy achieved by the proposed scheme against the two main FL methods in SGINs mentioned above. We employ ResNet-18 to train the CIFAR-10 and MNIST datasets under a non-independent and identically distributed (non-IID) distribution. For illustrative purposes, we assume eight clusters on Earth, each with five clients, where the images of clients within the same cluster are selected from only three labels. The LEO satellite constellation is constructed using Systems Tool Kit (STK), based on the Starlink system. We set the training epochs to 200 for CIFAR-10 and 80 for MNIST for these three methods. The results show that the proposed scheme achieves the highest test accuracy within a relatively short training time. Furthermore, it avoids using feeder links or ISLs for model aggregation or exchange, thereby reducing communication overhead. A comparison of the communication costs between the proposed algorithm and existing approaches can be found in our previous work \cite{chen2024fedmeld}.

\subsection{Space-ground Fluid Inference}\label{subsec:inference}
Edge devices in remote regions can offload compute-intensive inference tasks to fluid AI. To achieve low E2E latency, the main challenges come from the limited computing capabilities of satellite nodes and the ever-changing network topology, making judicious service routing and migration among these orbiting servers a must. In this regard, we will discuss the second component of fluid AI, called fluid inference, to support in-network inference, for which inference accuracy is progressively increased via propagation through model layers distributed at satellites or ground stations.

\begin{figure}[h]
	\centering	\includegraphics[width=0.48\textwidth]{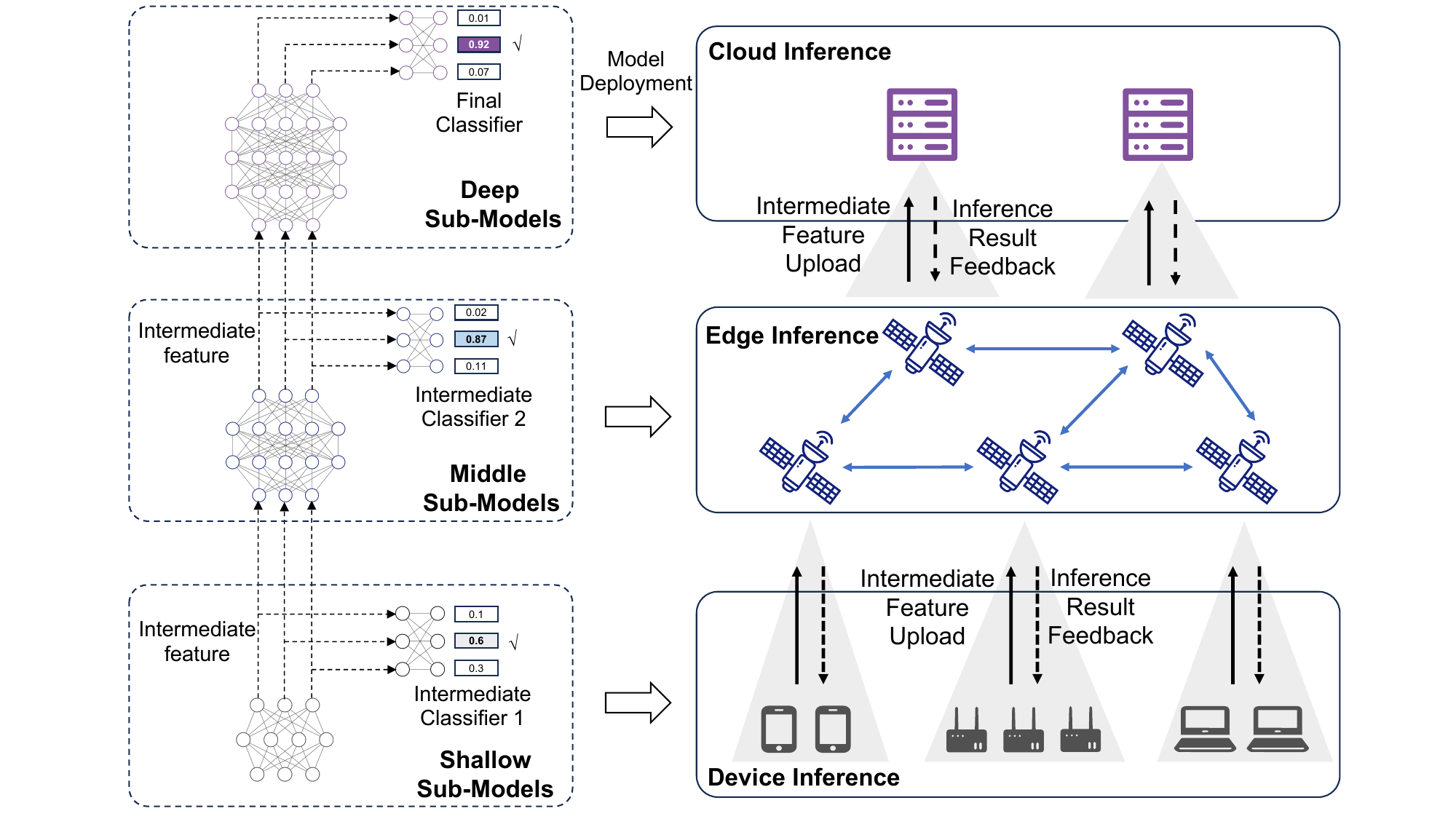}
	\caption{Cascading inference framework in fluid AI systems. This adaptive progressive inference protocol can stop early as long as the accuracy of the intermediate classification result satisfies the user's QoS requirements.}
	\label{fig:progressive_inference}
\end{figure}

Considering the varying computing capacities of end-edge-cloud systems within SGINs, we propose partitioning the neural network into multiple cascading sub-models, categorized as head, middle, and tail sub-models, as illustrated in Fig. \ref{fig:progressive_inference}. The head sub-model can extract low-level features with shallow neural layers, making it ideal for deployment on clients with limited computing resources. The middle sub-model handles several middle layers with moderate computing complexity, which is suitable for deployment on satellite nodes. The tail sub-model often contains the majority of neural layers responsible for high-level feature extraction and complex logical operations, and can be deployed at ground data centers with abundant computing resources. Depending on resource availability and communication link capacity, this three-tier architecture can be adapted to a two-tier framework, such as user–satellite cooperation, or expanded to multi-tier frameworks, such as propagation through multiple satellite nodes. With early exiting techniques, each sub-model is equipped with a low-complexity intermediate classifier capable of generating intermediate inference results based on the sub-model, thereby maintaining a balance between accuracy and total inference latency.

To reduce E2E latency, the inference process can be terminated early at any sub-model using an intermediate classifier, thus avoiding the need to complete the entire model computation. However, this comes at the cost of reduced inference accuracy \cite{9955582}. As a result, inference tasks can propagate different nodes according to their quality of service (QoS) requirements. Considering task request probability from the ground, model placement and routing can be studied in fluid AI systems. Since the cascaded model placement remains unchanged for a certain period, the average inference accuracy/latency during this time window is determined by the satellites’ mobility patterns. For this reason, the mobility of satellite nodes greatly complicates the accuracy-latency tradeoff in space–ground fluid inference, distinguishing the design from its terrestrial counterpart.

Another key design issue is to formulate an inference task (model) migration problem. More specifically, to maximize the task completion rate, both horizontal and vertical task migration can be conducted in a fluid-like manner—that is, migrating models either over inter-satellite links or between ground stations and satellites. The former aims to balance the communication-computing workload across satellites, while the latter dynamically controls the inference offloading to either distant cloud centers or nearby visible satellites by comparing offloading latency and migration costs.

\subsection{Space-ground Fluid Model Downloading}\label{subsec:download}
The aforementioned learning and inference components cannot be made possible without effective model delivery over fluid AI. To fulfill this vision, we design the final component of fluid AI, called fluid model downloading, to enable efficient model downloading to edge devices via fluid AI. Given the capacity-limited feeder links and a massive number of ground users, fluid AI should enable efficient model downloading by fully utilizing the limited spectrum bandwidth. To improve caching and spectrum efficiency, our core idea is to explore knowledge reuse, which is a recent trend in multi-task learning and parameter-efficient fine-tuning in which parameter blocks can be shared across deep neural networks for versatile tasks~\cite{10471279}. The details are elaborated below.

\subsubsection{Parameter-sharing model caching and migration}

\begin{figure}[h]
	\centering
	\includegraphics[width=0.48\textwidth]{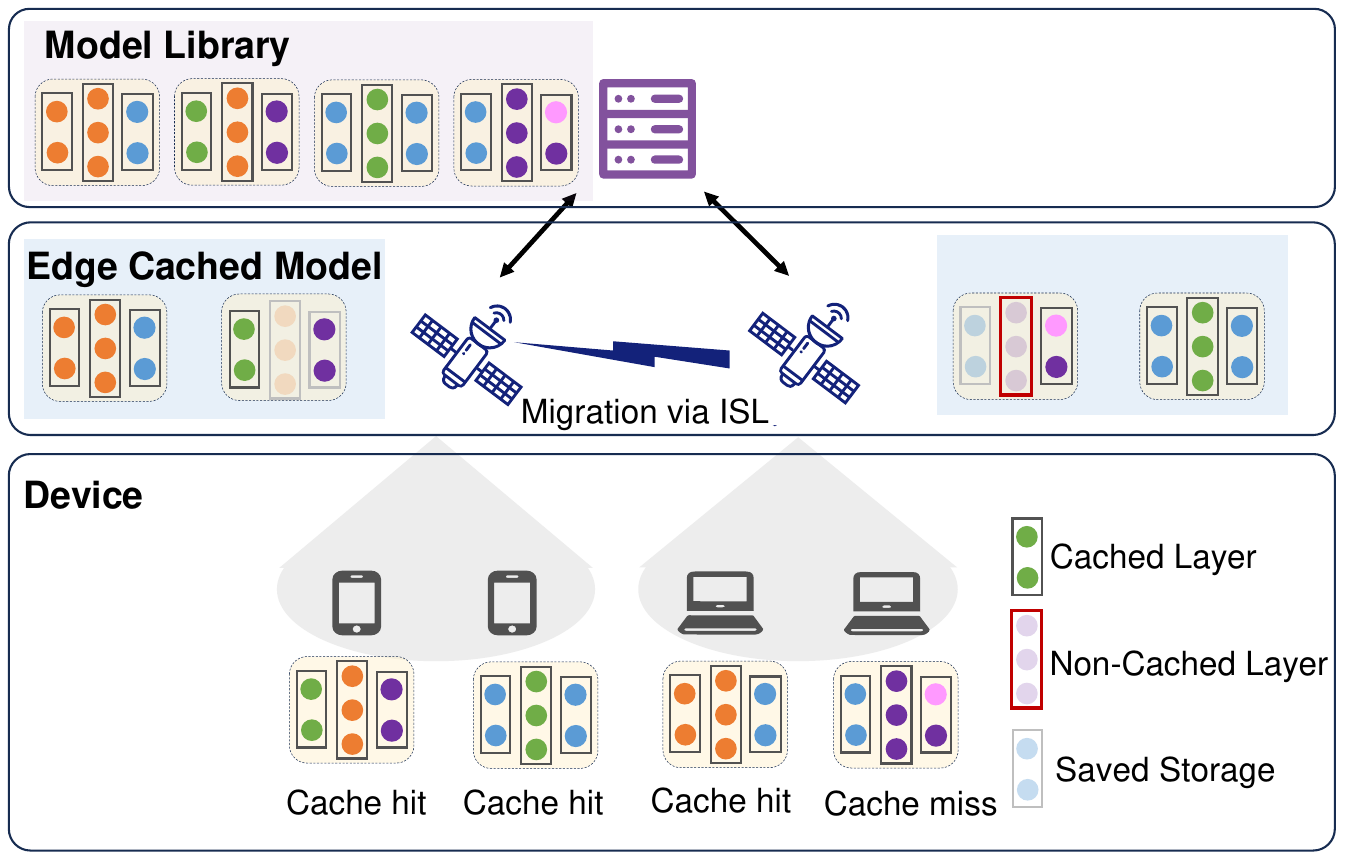}
	\caption{Parameter-sharing caching scheme for fluid AI systems.}
	\label{fig:illustration_layer_sharing}
\end{figure}

The first problem is where to cache the model in fluid AI networks to facilitate ultra-fast model downloading. In parameter-efficient fine-tuning, layer-freezing transfer learning, or multi-task learning, a significant proportion of parameters in deep neural networks can be shared among different tasks because they contain foundational knowledge applicable across various downstream applications~\cite{qu2024trimcaching}. Therefore, when multiple AI models with shared parameters are stored on the same satellite, the common parameter can be shared among these models, improving storage efficiency. Building on this concept, rather than caching all the parameters of each model on the satellites, we can allow each satellite to cache only certain parameter blocks of different models, while neighboring satellites can migrate some layer parameters via laser ISLs when necessary, as illustrated in Fig. \ref{fig:illustration_layer_sharing}.
This parameter-sharing caching mechanism efficiently leverages limited storage resources to increase the cache hit ratio of edge satellites when serving ground users. By retrieving frequently requested model components from visible satellites rather than a remote cloud center, our scheme effectively reduces model downloading latency. To maximize the cache hit ratio, an optimal scheme must be devised to meet users’ accuracy and latency requirements. The mobility patterns of satellite nodes and the hierarchical space–ground architecture distinguish the optimal caching scheme from its terrestrial counterparts, such as in our previous work~\cite{qu2024trimcaching}.

\subsubsection{Parameter-sharing for AI downloading}
When supporting the simultaneous downloading of diverse and high-dimensional models for a large number of devices, space-to-ground wireless links pose a major communication bottleneck. The traditional unicasting scheme requires each task to download a complete model individually. However, a single satellite must serve users in a large geographical area simultaneously, causing users to experience excessively long wait times during peak periods. This situation is exaggerated by the limited bandwidth and transmit power, as well as unreliable transmission links (e.g., caused by adverse weather conditions), in satellite networks.

To fully utilize knowledge reuse for communication-efficient model downloading across multiple devices, parameter-sharing multicasting can be adopted~\cite{10471279}. Specifically, a satellite multicasts model parameters (blocks) that are reusable across different tasks and devices, allowing multiple devices to download these parameters simultaneously. In this context, careful radio resource management is essential, including beam footprint adaptation, power control, and bandwidth allocation, to optimize E2E performance and maximize the efficiency of AI model downloading. Furthermore, joint design of satellite model caching and multicasting is essential to achieve ultimate performance by considering the coordination among satellite nodes with diverse trajectories.

\section{Deployment Considerations and Future Directions of Space-Ground Fluid AI}\label{sec:deploy_future}

\subsection{Deployment Considerations of Fluid AI}
The deployment of fluid AI onboard satellites must account for two fundamental challenges posed by the space environment: the harsh physical conditions of space and the intermittent nature of satellite power supply. These factors significantly influence the feasibility and reliability of AI service provisioning in SGINs. Unlike terrestrial AI systems, satellites operate in an environment characterized by high levels of radiation, extreme temperature fluctuations, and a vacuum atmosphere. Studies have shown that failures induced by space radiation account for approximately 40\% of total satellite failures, highlighting the severity of this issue. Radiation exposure can lead to hardware degradation and bit-flip errors in AI processors, affecting model accuracy and stability over time. Moreover, temperature variations between sunlit and shadowed regions can cause thermal expansion and contraction, potentially leading to hardware stress and increased failure rates. To address these issues, satellites are designed with radiation-hardened electronic components and error-correcting mechanisms, which increase system resilience against cosmic radiation. Fault-tolerant computing strategies, such as triple modular redundancy (TMR), also play a crucial role in mitigating computation errors by running the same process on three separate units and using majority voting to determine the correct output. Additionally, redundant AI model storage ensures that, even if a model is corrupted due to radiation-induced faults, a backup version can be retrieved from non-volatile memory, enabling continuous AI inference and learning. From a networking perspective, random graph theory can be employed to address computing power deployment and service (model) placement on SGINs, thereby ensuring that time-sensitive tasks can be forwarded to other neighboring satellite nodes under node failure and maintaining high reliability for mission-critical services.
Another major constraint in deploying AI on satellites is the limited and variable power supply. Unlike terrestrial infrastructure, which has stable power sources, satellites rely primarily on solar panels to generate energy. In the sunlit region, a satellite can harvest and store energy, but once it enters the shadowed region, power generation is suspended, making it necessary to operate on stored battery reserves. Given that satellite orbits and their positions relative to the Sun are highly predictable, this characteristic can be exploited to implement energy-aware AI task scheduling. More specifically, inter-satellite workload balancing can optimize energy utilization by dynamically offloading tasks to satellites with surplus energy, reducing the likelihood of power depletion and ensuring uninterrupted AI operations.

\subsection{Future Directions of Fluid AI}
Fluid AI remains a largely unexplored area that presents a wealth of research opportunities. Several potential directions are listed below.

\begin{itemize}
    \item \textbf{Energy-efficient fluid AI.} Ground devices can increase transmit power to overcome the severe free-space path loss caused by long transmission distances. Thus, these ground devices will have higher energy consumption than cellular users, which poses significant challenges for the training process over a long time. Based on the predictable channel conditions, this issue can be alleviated by scheduling clients to upload information only when close to the satellite. However, doing so will lead to a waste of limited visible time between satellites and ground users. The tradeoff between energy consumption and time duration can be investigated to realize energy-efficient fluid AI.

    \item \textbf{Low-latency fluid AI.} High propagation delay (i.e., several milliseconds) caused by long transmission distances in SGINs is a primary bottleneck in supporting time-sensitive applications, such as autonomous driving. While signal exchanges between space and ground (i.e., acknowledgment information) are essential for reliable transmissions and high inference/training accuracy, such a procedure also causes excessive propagation delay. Thus, to reduce inference latency, satellite–ground signaling mechanisms must be carefully designed to strike the optimal tradeoff between inference latency and accuracy.

    \item \textbf{Secure fluid AI.} Due to the mobility of satellites, security and privacy issues remain largely underexplored in fluid AI. Satellites continuously traverse different regions, exposing themselves to a wider attack surface. Eavesdropping and interception threats are more severe in satellites, since adversaries can exploit long-range satellite links to extract sensitive model parameters or inference results. Additionally, FL across orbital and ground nodes increases the risk of model poisoning and backdoor attacks, in which malicious updates are injected from compromised ground nodes. The constrained computing resources of satellites also limit the implementation of complex defense mechanisms, compared with their terrestrial counterparts. However, the repetitive trajectories of satellite constellations offer a unique advantage. Security-enhanced models can be transferred from former satellites to later ones using techniques such as knowledge distillation, thereby reducing vulnerability to evolving threats and increasing system resilience.
\end{itemize}	

\section{Conclusion Remarks}\label{sec:conclusion}
In this article, we introduced fluid AI technology to explore the pivotal role of AI within SGINs. Unlike terrestrial edge AI, fluid AI leverages the predictable trajectories of satellites to inform AI service provisioning. Considering SGINs’ periodic network topologies, we have comprehensively reviewed the architecture, applications, and unique advantages that SGINs bring to edge AI. To address the challenges of mobility management and space–ground network cooperation, we have also outlined enabling techniques for efficient fluid learning, inference, and model downloading within SGINs. Fluid AI represents a pioneering step toward seamlessly integrating edge AI and SGINs in the upcoming 6G era. We hope this article will inspire further research into harnessing the potential of SGINs for advancing efficient edge intelligence strategies.

	\ifCLASSOPTIONcaptionsoff
	\newpage
	\fi

	\bibliographystyle{IEEEtran}
	\bibliography{IEEEabrv,body/reference}
	
	%
	%
	%
	%
	%
	%
	%
	%
	%
	%
	%
	%

\end{document}